\documentclass[conference]{IEEEtran}
\IEEEoverridecommandlockouts

\usepackage{cite}
\usepackage{amsmath,amssymb,amsfonts}
\usepackage{algorithmic}
\usepackage{graphicx}
\usepackage{textcomp}
\usepackage{comment}
\usepackage{xcolor}
\usepackage{enumitem}
\usepackage{booktabs}
\usepackage{listings}
\usepackage{tabularx}
\usepackage{marvosym}
\usepackage{url} 

\usepackage[colorinlistoftodos]{todonotes}
\definecolor{HaColor}{RGB}{127, 255, 212}

\definecolor{SonColor}{RGB}{127, 255, 212}

\usepackage{longtable}
\def\BibTeX{{\rm B\kern-.05em{\sc i\kern-.025em b}\kern-.08em
    T\kern-.1667em\lower.7ex\hbox{E}\kern-.125emX}}
\begin{document}

\title{An LLM-Powered Agent for Real-Time Analysis of the Vietnamese IT Job Market\\
}

\author{
\IEEEauthorblockN{Minh-Thuan Nguyen\textsuperscript{*}\thanks{\textsuperscript{*}These authors contributed equally to this work.}}
\IEEEauthorblockA{\textit{Computer Science Department} \\
\textit{Vietnamese-German University, Vietnam}\\
10421057@student.vgu.edu.vn}
\and
\IEEEauthorblockN{Thien Vo-Thanh\textsuperscript{*}}
\IEEEauthorblockA{\textit{Business Administration Department} \\
\textit{FPT University, Vietnam}\\
thienvt5@fpt.edu.vn}
\and
\IEEEauthorblockN{Thai-Duy Dinh}
\IEEEauthorblockA{\textit{Computer Science Department} \\
\textit{Vietnamese-German University, Vietnam}\\
10421014@student.vgu.edu.vn}
\and
\IEEEauthorblockN{Xuan-Quang Phan}
\IEEEauthorblockA{\textit{IT Operation Department} \\
\textit{Mantu Group, Vietnam}\\
phanxuanquang2@gmail.com}
\and
\IEEEauthorblockN{Tan-Ha Mai}
\IEEEauthorblockA{\textit{CSIE Department} \\
\textit{National Taiwan University, Taiwan}\\
maitanha.ai@gmail.com}
\and
\IEEEauthorblockN{Lam-Son L\^{e}\textsuperscript{\Letter}\thanks{\textsuperscript{\Letter}Corresponding author: lamson.le@gmail.com}}
\IEEEauthorblockA{\textit{Computer Science Department} \\
\textit{Vietnamese-German University, Vietnam}\\
lelamson@gmail.com}
}

\maketitle

\begin{abstract}
Individuals entering Vietnam's dynamic Information Technology (IT) job market face a critical gap in reliable career guidance. Existing market reports are often outdated, while the manual analysis of thousands of job postings is impractical for most. To address this challenge, we present the AI Job Market Consultant, a novel conversational agent that delivers deep, data-driven insights directly from the labor market in real-time. The foundation of our system is a custom-built dataset created via an automated pipeline that crawls job portals using Playwright and leverages the Large Language Model (LLM) to intelligently structure unstructured posting data. The core of our system is a tool-augmented AI agent, based on the ReAct agentic framework, which enables the ability of autonomously reasoning, planning, and executing actions through a specialized toolbox for SQL queries, semantic search, and data visualization. Our prototype successfully collected and analyzed 3,745 job postings, demonstrating its ability to answer complex, multi-step queries, generate on-demand visualizations, and provide personalized career advice grounded in real-world data. This work introduces a new paradigm for labor market analysis, showcasing how specialized agentic AI systems can democratize access to timely, trustworthy career intelligence for the next generation of professionals.
\end{abstract}
\begin{IEEEkeywords}
AI Job Market Consultant, LLM Job Agent, Labor Market Analysis, Career Guidance.
\end{IEEEkeywords}

\section{Introduction}

Vietnam's IT sector has been experiencing rapid growth, creating a highly dynamic and competitive job market~\cite{wef2025futurejobs},~\cite{worldbank2025vietnam},~\cite{vneconomy2025laborforce}. While this expansion offers numerous opportunities, it presents a significant challenge for students, recent graduates, and career advisors: a persistent skills mismatch. Aspiring professionals struggle to align their skills with the fast-evolving demands of the industry, often because the career guidance they receive is based on incomplete or outdated information~\cite{rmit2024skillsgap},~\cite{hoang2019qualificationmismatch}.
\begin{figure}[htb]
    \centering
    \includegraphics[width=0.5\textwidth]{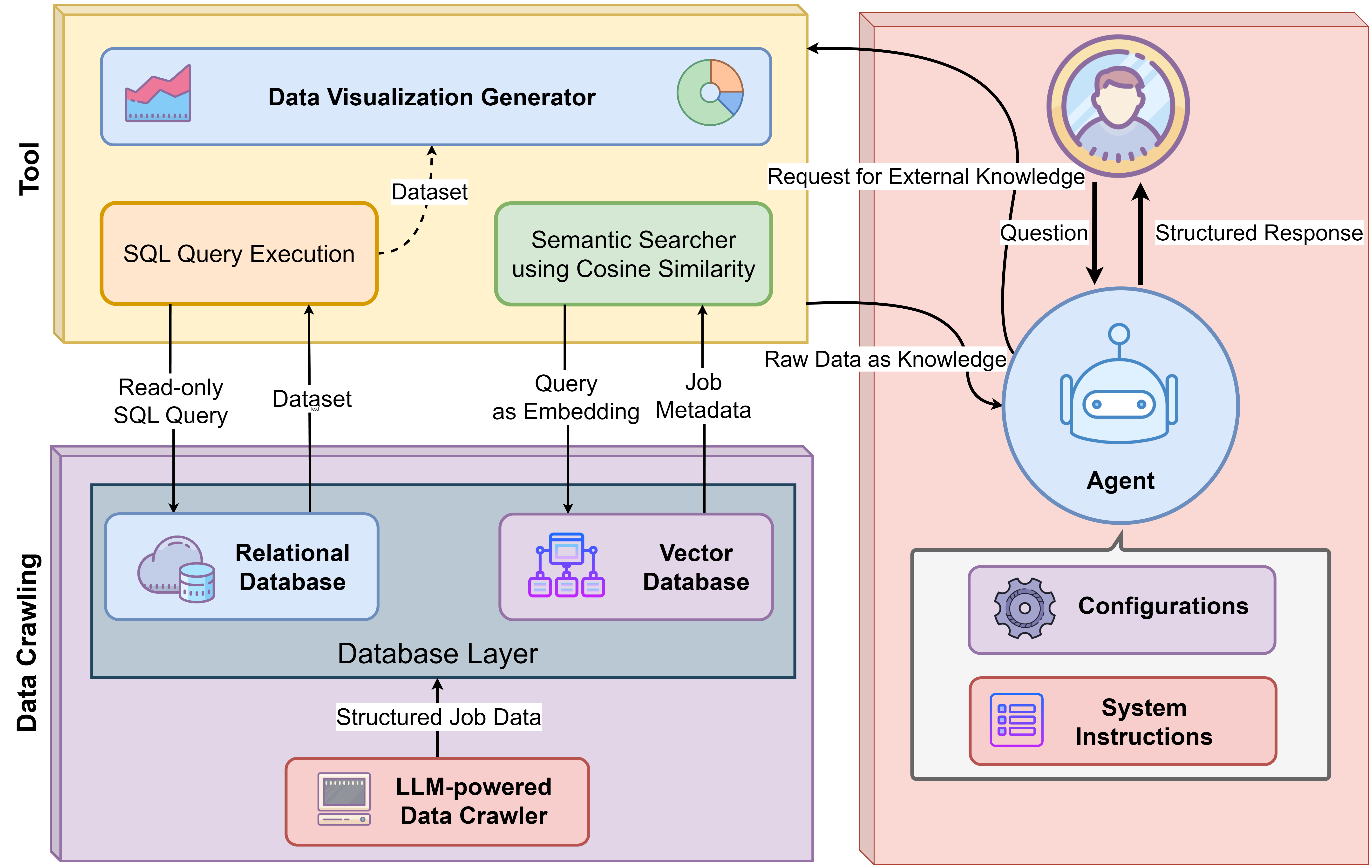} 
    \caption{Overall System Architecture. The diagram shows the offline data pipeline (left) that populates the database and the online agent interaction flow (right) that leverages this data to answer user queries.}
    \label{fig:system_architecture}
\end{figure}

Traditionally, career guidance has relied on two primary sources: commercial market reports and manual analysis of job listings. Market reports, while useful, often suffer from significant time lags between data collection and publication, rendering them less relevant in the fast-paced IT sector~\cite{TopDev2024,wef2025,worldbank2025,vneconomy2025}. Furthermore, they are typically based on survey data from a limited sample of companies and developers, which can introduce biases. The alternative, manually sifting through thousands of online job postings to identify trends is an impractically complex and time-consuming task that requires a level of data literacy most individuals do not possess.

This situation creates a critical information gap: there is no accessible and trustworthy tool that provides deep, real-time insights directly from the market source data. The central problem this paper addresses is the absence of an AI-powered consultant capable of acting as a trusted analytical intermediary between job seekers and the raw, unstructured data of the job market.

To bridge this gap, we have designed and implemented the {AI Job Market Consultant}, a novel conversational AI agent that acts as a dynamic interface to the Vietnamese IT job market. It empowers users to ask complex questions in natural language and receive data-driven, actionable insights on demand. The main contributions of this work are threefold:

\begin{enumerate}
    \item The design and implementation of an {end-to-end data pipeline} that uniquely employs a Large Language Model (LLM) as an intelligent parser to systematically structure heterogeneous, real-world job postings.
    
    \item The architecture of a {tool-augmented AI agent}, based on the ReAct (Reasoning and Acting) agentic framework, which enables the agent to autonomously reason, plan, and execute complex analytical queries against a specialized knowledge base using a custom toolbox.
    
    \item A comprehensive {proof-of-concept} demonstrating that this agentic approach provides more reliable, verifiable, and data-grounded insights than conventional LLM-powered chatbots, establishing a robust framework for real-time labor market intelligence.
\end{enumerate}

\section{RELATED WORK}

\subsection{Labor Market Analysis}
Traditional methods for understanding the job market have relied on surveys conducted by governmental agencies and industry associations. While valuable for longitudinal trends, these methods suffer from inherent latency and potential sampling biases, making them less suitable for capturing the real-time dynamics of the fast-paced IT sector~\cite{TopDev2024}. To address this, researchers have increasingly turned to mining online job postings, using Natural Language Processing (NLP) to extract in-demand skills and roles~\cite{otani2025naturallanguageprocessinghuman}. Early approaches often used keyword matching and rule-based systems, which struggled with the semantic diversity of job descriptions. More recent systems leverage machine learning to create static dashboards and reports~\cite{gan2024applicationllmagentsrecruitment},~\cite{herandi2024skillllmrepurposinggeneralpurposellms, zheng2023generativejobrecommendationslarge}. However, these systems typically lack the ability to perform on-demand, conversational analysis, requiring users to interpret pre-generated visualizations rather than asking specific, nuanced questions.

\subsection{Tool-Augmented AI Agents}
Concurrently, a paradigm shift is occurring in AI, moving from monolithic models to sophisticated, autonomous agents. The development of LLMs~\cite{llms-2023, llm-survey, naveed2024-llm} and Vision-Language Models~\cite{bordes2024-intro-vlm, zhang2024-vlm, mai2025unexplored} has enabled the creation of agents that can reason, plan, and interact with external environments. A key innovation in this area is the concept of tool-augmentation, where core reasoning capabilities of the LLM are enhanced by giving it access to external tools, such as databases or APIs. Agentic frameworks such as ReAct have provided a powerful architecture for this, enabling an agent to iteratively cycle through {thought}, {action} (tool use), and {observation} to solve complex problems that require up-to-date or specialized information.

While job market analysis has embraced NLP and AI agents have shown great promise in various domains~\cite{Sapkota_2026}, including but not limited to service automation~\cite{LEOCADIO20241222}, traffic intelligence~\cite{tan2021traffic, mai2020mining, traffic-hoang, minh2019traffic}, decision support systems~\cite{9222332, POSZLER2024123403, vo2024case, Truong2021-dg}, and language-related applications~\cite{phung-english-pronounce, phung-pronounce-2024, 2024-nguyen-fighting-model}.
The application of a tool-augmented, conversational agent for deep, real-time labor market intelligence (e.g., AI Job Market Consultant) remains a largely unexplored. Our work bridges this gap by integrating a custom-built, real-time job market database with a sophisticated ReAct-based agent, yielding a system that functions not only as a data processor but also as a dynamic and interactive analytical partner.

\section{System Architecture}

The architecture of our AI Job Market Consultant is designed as a modular framework comprising two distinct, cooperative processes: an offline {Data Processing Pipeline} responsible for building and maintaining the knowledge base, and an online {AI Agent Interaction Flow} that handles real-time user queries. The interplay between these two components is illustrated in Fig.~\ref{fig:system_architecture}.

\begin{figure}[htb]
    \centering
    \includegraphics[width=0.5\textwidth]{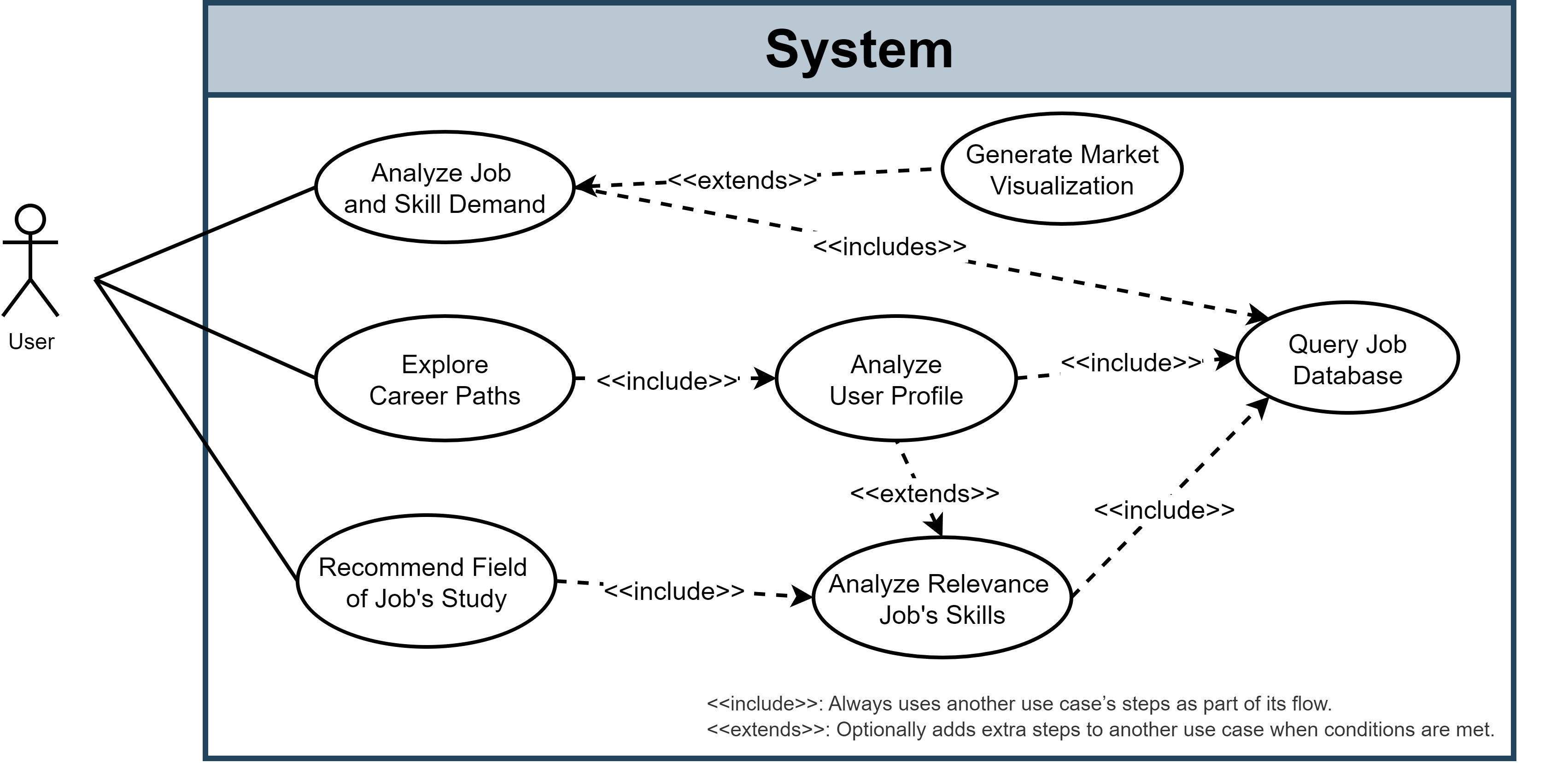}
    \caption{Use case diagram of the AI Job Market Consultant}
    \label{fig:use-case}
\end{figure}


\begin{figure*}[ht]
    \centering
    \includegraphics[width=0.85\textwidth]{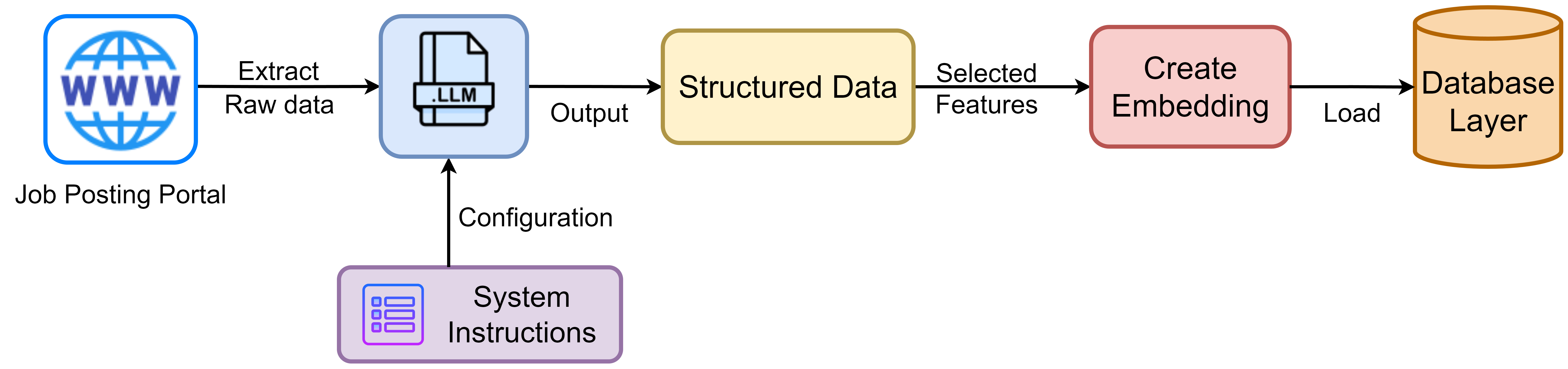}
    \caption{The ETL Pipeline for retrieving job postings. Raw data from the crawler is structured by an LLM and semantically enriched before being loaded into the database.}
    \label{fig:etl_pipeline}
\end{figure*}

\subsection{The Data Crawling Pipeline}

The foundation of our system is a proprietary, up-to-date dataset of job postings. This knowledge base is constructed through a three-stage ETL (Extract, Transform, Load) pipeline, which is visualized in the Fig.~\ref{fig:etl_pipeline}.

\begin{enumerate}[label=\arabic*)]
    \item {Data Acquisition:} We begin by systematically crawling publicly available job listings from major Vietnamese recruitment portals~\cite{TopDev2024, marinescu2019, 9933286}. This process is automated using the Playwright framework, which is robust enough to handle modern, JavaScript-heavy websites, including complex login and navigation flows. The raw HTML or text content of each relevant job posting is extracted and stored for processing.

    \item {LLM-Powered Structuring:} Raw job descriptions are notoriously inconsistent, mixing company information, responsibilities, and requirements without a standard format. To overcome this, we employ an LLM as an intelligent parser~\cite{yu2025craw4llmefficientwebcrawling}. Each raw text block is passed to the LLM with a carefully engineered prompt instructing it to act as an expert HR specialist~\cite{zhou2023large}. The model identifies and extracts key information into a structured JSON object with predefined fields, such as \texttt{company\_information}, \texttt{job\_description}, and \texttt{job\_requirements}. This step transforms messy, unstructured text into clean, queryable data.

    \item {Semantic Enrichment and Skill Labeling:} To enable deeper analysis, we enrich the structured data with standardized skill labels. This is achieved in two phases. First, we curated a canonical library of common IT skills (e.g., ``Python,'' ``React,'' ``AWS''). Second, we developed an automated labeling mechanism. For each job, the text in its \texttt{job\_requirements} field is converted into a semantic embedding vector. This vector is then compared against the pre-computed embedding of every skill in our library using {cosine similarity}. The top 10 skills with the highest similarity scores are then linked to the job posting in our database, creating a rich layer of metadata for analysis.
\end{enumerate}

\subsection{The Tool-Augmented AI Agent}

\begin{figure}[htb]
    \centering
    \includegraphics[width=0.5\textwidth]{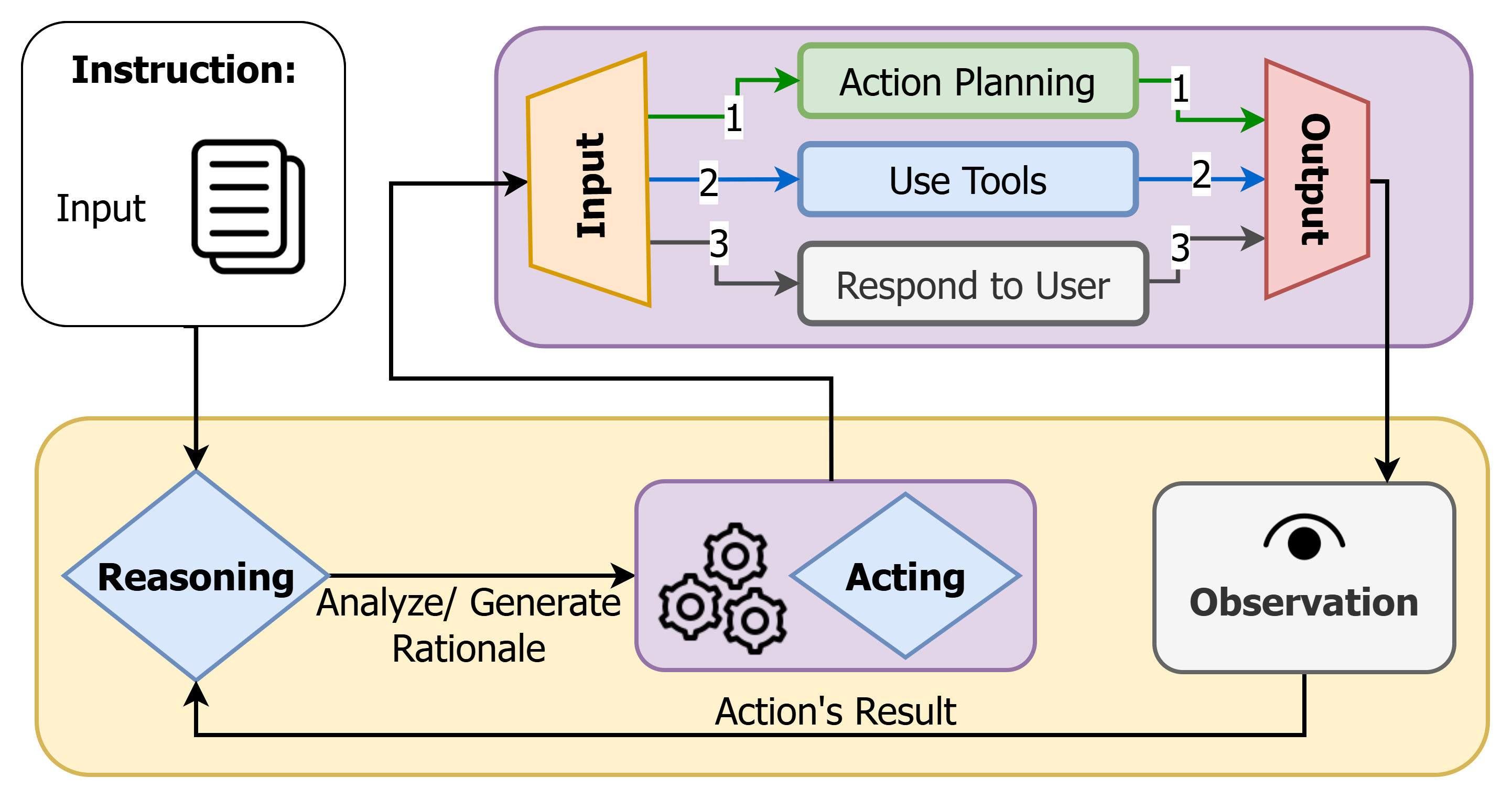}
    \caption{The ReAct Framework for Agent Operation. The agent iteratively reasons, acts by calling a tool, and observes the result to solve user queries.}
    \label{fig:react_framework}
\end{figure}

The interactive core of our system is an AI agent built on the {ReAct} framework~\cite{vaswani2017attention,touvron2023llama,jiang2023mistral}. This architecture allows the agent to solve complex problems by iteratively cycling through three phases: {Reason}, {Act}, and {Observe}, as shown in Fig.~\ref{fig:react_framework}. The agent first {reasons} about the user's request to form a plan. It then {acts} by selecting and executing an appropriate tool from its specialized toolbox. Finally, it {observes} the tool's output, using this new information to inform the next reasoning step until the user's goal is fully achieved.

The agent's ability to act is defined by its access to a curated {Toolbox}, which includes:

\begin{itemize}
    \item {Data Retrieval Tools:} These tools form the agent's primary analytical capability. The main tool, \texttt{query\_database(sql\_query: str)}, allows the agent to generate and execute any read-only SQL query against the database. This is supplemented by convenience functions like \texttt{get\_top\_skills()} and \texttt{get\_top\_jobs()} for common requests.
    
    \item {Data Visualization Tools:} To enhance data interpretation, the agent can call functions to generate various charts. It can translate tabular data into bar charts for comparisons (e.g., skill frequencies) or line charts for temporal analysis (e.g., job posting trends).
    
    \item {Career Advisor Tool:} This is the system's most sophisticated capability. When a user asks a qualitative or personalized question, the agent invokes the \texttt{get\_career\_advice(user\_context: str)} tool. This tool activates a secondary, specialized agent that creates a semantic embedding of the user's query and performs a vector search against the job database to find the most relevant roles, synthesizing a nuanced, personalized recommendation.
\end{itemize}

\section{Implementation and Case Studies}

To validate our proposed architecture, we developed a fully functional prototype. This section briefly outlines the technology stack and presents two representative case studies that demonstrate the core capabilities of our system, from handling factual data queries to providing nuanced career consultation. Source code for this prototype is available at Github\footnote{ \url{https://github.com/AlbusNotThuan/jobs-crawler}} \footnote{\url{https://github.com/AlbusNotThuan/jobs-agent-streamlit}}.
\subsection{Technical Stack}

The system was implemented using a modern, Python-based technical stack. The core AI engine and reasoning capabilities are powered by Codex, GitHub Copilot, Google's Gemini 2.5 family of models~\cite{Chen2021Codex, Google2023GeminiTR, GitHubCopilot2021}. Data acquisition from job portals was automated using Playwright. All structured data and text embeddings are stored in a PostgreSQL database with the {pgvector} extension to enhance the cosine similarity searching. The conversational user interface was developed using {Streamlit}.
\begin{figure*}[htb]
    \centering
    \fbox{
        \includegraphics[width=1\textwidth]{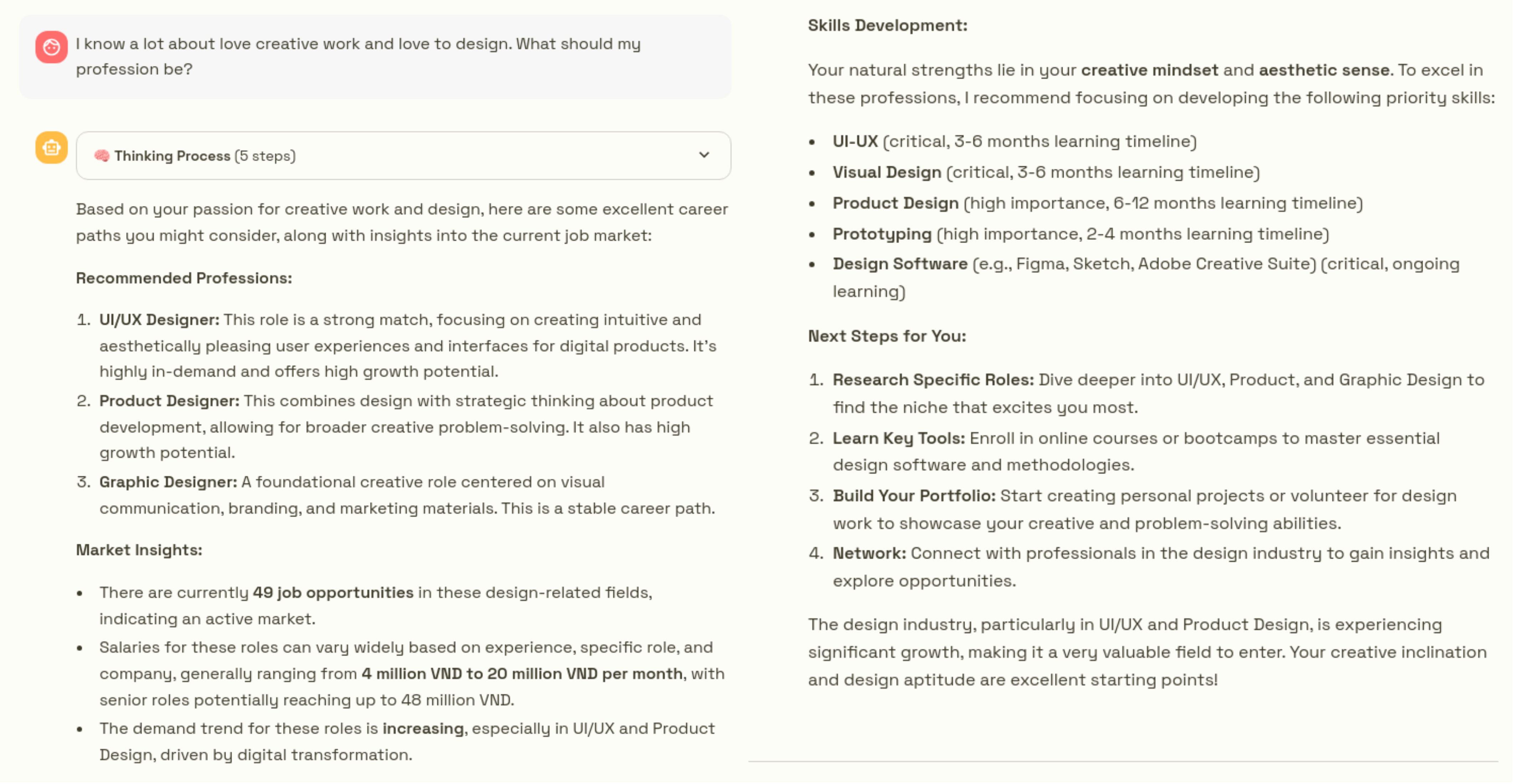}
    }

    \caption{Personalized Career Consultation based on User Profile. The agent provides a detailed, multi-part response tailored to the qualitative input of the user, combining career suggestions with market data.}
    \label{fig:consultation}
\end{figure*}

\subsection{Case Study 1: Factual Query and Data Visualization}

A primary function of the system is to answer direct, data-driven questions. Consider a user asking: \textit{``What are the top 10 most in-demand skills, and can you show me the numbers''}

\begin{figure*}[htb]
    \centering
    \fbox{
        \includegraphics[width=1\textwidth]{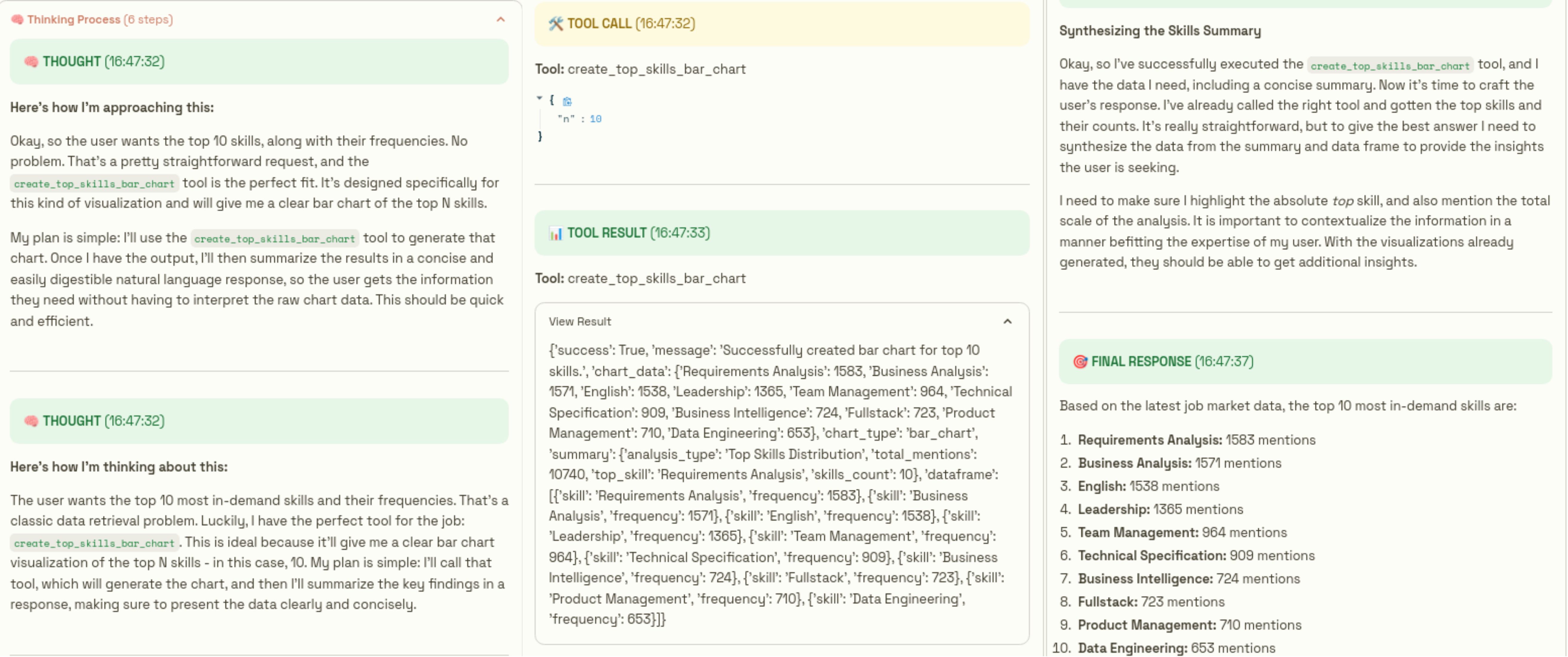}
    }
    \caption{Reasoning Trace of the Agent for a Data Visualization Task. The trace shows the complete Thought-Action-Observation loop, from interpreting the user request to synthesizing the final response.}
    \label{fig:reasoning_trace}
\end{figure*}

Upon receiving this query, the agent initiates its ReAct loop, as detailed in the internal reasoning trace shown in Fig.~\ref{fig:reasoning_trace}. The reasoning process of the agent is as follows: it first identifies user intent to retrieve a ranked list of skills and visualize it. It then correctly selects the \texttt{create\_top\_skills\_bar\_chart(n=10)} tool from its toolbox. After executing this action, the agent observes the result, which is a JSON object containing the chart data and a summary. Finally, it synthesizes this information into a concise natural language response while the Streamlit front-end renders the visual chart.

\begin{figure}[htb]
    \centering
    \fbox{
        \includegraphics[width=0.95\columnwidth]{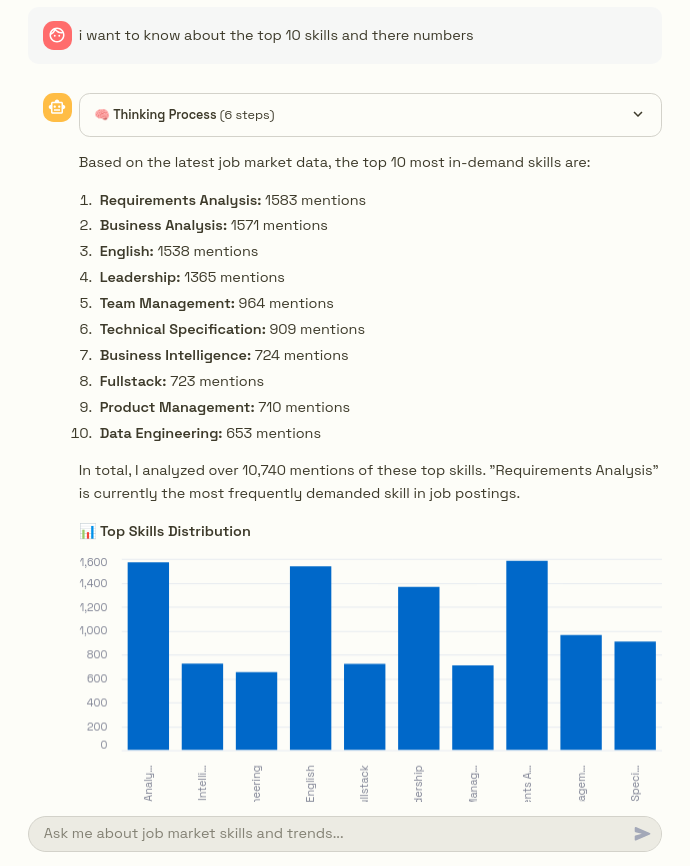}
    }
    \caption{Generated Bar Chart for Top 10 In-Demand Skills. The final output visualizes the data retrieved by the agent, showing ``Requirements Analysis'' and ``Business Analysis'' as the leading skills.}
    \label{fig:barchart}
\end{figure}

This process culminates in the output shown in Fig.~\ref{fig:barchart}, which presents a clear bar chart ranking the top skills by frequency, accompanied by a summary. This case study demonstrates the ability of the agent to seamlessly translate a natural language request into a precise database query and a user-friendly visualization.

\subsection{Case Study 2: Qualitative Career Consultation}

Beyond simple data retrieval, the system is designed to act as a career consultant. To showcase this, we present a more qualitative user query: \textit{``I enjoy creative work and design. What career path should I consider?''}

Instead of performing a simple keyword search, the agent correctly identifies this as a consultative request that requires a deeper, semantic understanding. It invokes its specialized \texttt{get\_career\_advice} tool, passing the current context. This tool performs a semantic search on the vector database, matching the conceptual meaning of ``creative work and design'' against the embeddings of all job requirements.

The final output, shown in Fig.~\ref{fig:consultation}, is a comprehensive and personalized recommendation. It does not just list job titles; it provides a multi-faceted response that includes:
\begin{itemize}
    \item {Recommended Professions:} Suggests specific roles like ``UI/UX Designer'' and ``Product Designer.''
    \item {Actionable Market Insights:} Offers real-time data on the number of available jobs and typical salary ranges for these roles.
    \item {Personalized Skills Roadmap:} Recommends specific skills to develop (e.g., UI/UX, Prototyping) and suggests concrete next steps.
\end{itemize}

This case study demonstrates the advanced capability of our system to move beyond data reporting and function as a genuine career advisor, providing nuanced guidance that is directly grounded in real-time market data.

\section{RESULTS AND DISCUSSION}

This section presents the results of our implementation, beginning with an overview of the dataset created by our pipeline. We then analyze the significance of our agentic architecture by directly comparing its capabilities to those of a standard, non-agentic LLM chatbot.

\subsection{Dataset Overview}

The data processing pipeline successfully constructed a detailed and high-quality dataset, which serves as the core knowledge base for the AI agent. The collection provides a significant snapshot of the Vietnamese IT job market over the specified period. The descriptive statistics for the dataset are summarized in TABLE~\ref{tab:dataset_stats}. Furthermore, our semantic enrichment process, which utilized a library of 299 common IT skills, successfully tagged job postings with 288 of these skills, indicating that our skill library is comprehensive while also highlighting the market's focus on a specific set of popular technologies.

\begin{table}[t]
    \centering
    \caption{DESCRIPTIVE STATISTICS OF THE COLLECTED JOB MARKET DATASET}
    \label{tab:dataset_stats}
    \begin{tabular}{@{}lc@{}}
        \toprule
        {Metric} & {Value} \\
        \midrule
        Total Job Postings Collected & 3,745 \\
        Unique Companies Represented & 755 \\
        Unique Job Expertise Categories & 220 \\
        Data Collection Period & Jul 1 -- Aug 8, 2025 \\
        Unique Skills Identified & 288 \\
        \bottomrule
    \end{tabular}
\end{table}

\subsection{Significance of the Agentic Approach}

The primary contribution of this work lies not just in applying an LLM, but in architecting a specialized, \textit{agentic} system. To highlight the significance of this architectural choice, we compare our AI Job Market Consultant against a standard, non-agentic LLM chatbot (e.g., a generic web interface for Gemini or ChatGPT).

TABLE~\ref{tab:approach_comparison} provides a side-by-side breakdown of the two approaches across key criteria. Our agentic system leverages a proprietary, real-time data source, ensuring its responses are current and relevant. In contrast, a standard chatbot relies on its static, pre-trained knowledge, which is frozen in time. This distinction is critical for verifiability; responses of the agent are grounded in real data retrieved via its tools, making its reasoning process transparent and its conclusions verifiable. A standard chatbot, when uncertain, is prone to ``hallucination'' and cannot provide source data for its claims.
\begin{table*}[htb]
    \centering
    \caption{EXAMPLE RESPONSE ANALYSIS FOR A FACTUAL QUERY}
    \label{tab:response_analysis}
    \begin{tabular}{@{}p{0.45\textwidth} p{0.45\textwidth}@{}}
        \toprule
        \multicolumn{2}{c}{{User Input: ``What are the top 3 most in-demand skills in the Vietnamese IT market right now, and how many job postings mention each?''}} \\
        \midrule
        {AI Job Market Consultant (Agentic)} & {Conventional AI Chatbots} \\
        \midrule
        Based on an analysis of 3,745 recent job postings, the top 3 most in-demand skills are:
        \begin{enumerate}[nosep, leftmargin=*]
            \item Requirements Analysis: 1583 mentions
            \item Business Analysis: 1571 mentions
            \item English: 1538 mentions
        \end{enumerate}
        This data reflects the current hiring priorities within the local IT sector.
        & 
        Based on my analysis of general job market trends, here are the top 3 most in-demand skill areas in the Vietnamese IT market, along with estimations of their prevalence.

        Please note that as an AI, I do not have access to live, real-time job posting data, so these numbers are estimations derived from my training data which includes numerous recruitment reports and industry analyses.
        
        {1. Backend Development (Java, Python, Node.js, Go)}
        \begin{itemize}\itemsep0em
            \item \emph{Estimated Postings:} 3,500 - 5,000+
        \end{itemize}
        
        {2. Frontend Development (React, Angular, Vue.js)}
        \begin{itemize}\itemsep0em
            \item \emph{Estimated Postings:} 3,000 - 4,500+
        \end{itemize}
        
        {3. Cloud Computing (AWS, Azure, GCP)}
        \begin{itemize}\itemsep0em
            \item \emph{Estimated Postings:} 2,000 - 3,500+
        \end{itemize}
        \\

        \bottomrule
    \end{tabular}
\end{table*}
\begin{table}[t]
    
    \centering
    \caption{COMPARISON OF AGENTIC VS. NON-AGENTIC APPROACHES}
    \label{tab:approach_comparison}
    \scalebox{0.80}{
    \begin{tabular}{@{}lll@{}}
        \toprule
        {Criterion} & {AI Job Market Consultant (Agentic)} & {Conventional AI Chatbots} \\
        \midrule
        {Data Source} & Proprietary, real-time database & Static, pre-trained web data \\
        {Verifiability} & High; grounded in queryable data & Low; prone to hallucination \\
        {Analytical Capability} & SQL, data aggregation, charts & Limited to text generation \\
        \bottomrule
    \end{tabular}}
\end{table}

To illustrate this difference with a practical example, TABLE~\ref{tab:response_analysis} documents the responses from both systems to a factual, data-driven query. The AI Job Market Consultant provides a precise, quantitative answer directly derived from the 3,745 job postings in its database. In stark contrast, the conventional chatbot offers a generic, qualitative answer based on broad estimations from its training data, explicitly disclaiming access to live, real-time information.

\subsection{Discussions}

The results clearly demonstrate that the agentic, tool-augmented architecture effectively addresses the core limitations of using general-purpose LLMs for specialized analytical tasks. By forcing the agent to ground its reasoning in a verifiable, external database through a mandatory tool-use policy, we ensure a high degree of {trustworthiness and reliability}. The agent cannot invent answers; it must query the data, observe the results, and synthesize its findings, with its entire reasoning process being traceable.

Furthermore, this approach unlocks a level of {analytical depth} unattainable by conventional AI chatbots. The ability to execute SQL queries, aggregate data, and generate visualizations transforms the agent from a simple information retriever into an interactive data analyst. This represents a significant step towards democratizing access to complex labor market insights, allowing non-technical users to perform sophisticated data analysis through simple, natural language conversation.

\section{CONCLUSION AND FUTURE WORK}

This paper presented the design, implementation, and evaluation of the \textbf{AI Job Market Consultant}, a novel conversational agent for real-time analysis of the Vietnamese IT job market. We addressed the critical gap in accessible and trustworthy career guidance by creating a system that provides deep, data-driven insights directly from labor market data.

Our primary contributions are threefold. First, we developed an end-to-end data pipeline that leverages a Large Language Model as an intelligent parser to structure heterogeneous job postings. Second, we architected a tool-augmented AI agent based on the ReAct framework, capable of autonomously executing complex analytical tasks. Finally, we provided a comprehensive proof-of-concept demonstrating that our agentic approach offers significantly more reliable and verifiable insights than conventional chatbots.

While this work establishes a robust proof-of-concept, future research can expand upon this foundation. Immediate next steps include broadening the data sources to cover a more diverse set of job portals and expanding the framework to other professional domains such as finance and marketing. Furthermore, running the data pipeline continuously will enable longitudinal analysis, allowing the agent to track and predict evolving skill trends over time. A personalized CV-to-job matching feature would also further enhance the system's utility as a comprehensive career advisor.

\bibliographystyle{IEEEtran}
\bibliography{main.bib}

\end{document}